\documentclass{article}

\usepackage{authblk}
\usepackage[utf8]{inputenc}
\usepackage{main} 
\usepackage{microtype}
\usepackage{times}
\usepackage{latexsym}
\usepackage{algorithm}
\usepackage{algorithmic}
\usepackage{graphicx}
\usepackage{xcolor}
\definecolor{deepgreen}{RGB}{34,139,34}
\usepackage{amsmath}
\usepackage{amssymb}
\usepackage{amsfonts}
\usepackage{mathtools}
\usepackage{bm}

\usepackage{graphicx}
\usepackage{subcaption}
\usepackage{float}
\usepackage{wrapfig}

\usepackage{booktabs}
\usepackage{array}
\usepackage{multirow}
\usepackage{multicol}
\usepackage{makecell}
\usepackage{longtable}
\usepackage{tabularx}
\usepackage{colortbl}
\usepackage{arydshln}

\usepackage{color}
\usepackage{xcolor}
\definecolor{mydarkblue}{rgb}{0,0.08,0.45}
\definecolor{wkblue}{rgb}{0.2, 0.3, 0.6}
\definecolor{meta-color}{rgb}{0.5, 0.5, 0.5}
\definecolor{bgblue}{RGB}{245,243,253}
\definecolor{ttblue}{RGB}{91,194,224}
\definecolor{mybrown}{RGB}{128,64,0}
\definecolor{titlecolor}{HTML}{4c9cff}
\definecolor{coolblue3}{rgb}{0.91, 0.94, 0.98}
\definecolor{myblue}{rgb}{0.9, 0.1, 0.94}
\definecolor{mygreen}{rgb}{0.64, 0.56, 0.88}
\definecolor{myyellow}{rgb}{0.68, 0.6, 0.1}
\definecolor{fancygreen}{rgb}{0.33, 0.68, 0.20}
\definecolor{salmon}{rgb}{0.94, 0.52, 0.49}
\definecolor{tablegreen}{rgb}{0.82, 0.94, 0.75}
\definecolor{tableblue}{rgb}{0.81, 0.90, 0.94}
\definecolor{tablered}{rgb}{0.97, 0.85, 0.85}
\definecolor{tableorange}{rgb}{0.96, 0.85, 0.81}

\usepackage{enumitem}
\usepackage{footnote}
\usepackage{lipsum}
\usepackage{setspace}

\usepackage{geometry}
\usepackage{fancyhdr}
\usepackage{lscape}
\usepackage{rotating}

\usepackage{algorithm}
\usepackage{algorithmic}

\usepackage[most]{tcolorbox}
\usepackage[framemethod=tikz]{mdframed}
\usepackage{awesomebox}

\usepackage{natbib}
\usepackage{url}
\usepackage[colorlinks=true,linkcolor=mydarkblue,citecolor=mydarkblue,filecolor=mydarkblue,urlcolor=mydarkblue]{hyperref}

\usepackage{bbding}
\usepackage{imakeidx}
\makeindex
\usepackage[toc]{multitoc}
\usepackage[edges]{forest}
\usepackage[normalem]{ulem}
\usepackage{fontawesome5}
\usepackage{blindtext}
\usepackage{pgfplots}
\usepackage{pgfplotstable}
\pgfplotsset{compat=1.18}

\usepackage{listings}
\usepackage{listingsutf8}
\newcommand\JSONnumbervaluestyle{\color{blue}}
\newcommand\JSONstringvaluestyle{\color{red}}

\newif\ifcolonfoundonthisline

\makeatletter
\lstdefinestyle{json}
{
  showstringspaces    = false,
  keywords            = {false,true},
  alsoletter          = 0123456789.,
  morestring          = [s]{"}{"},
  stringstyle         = \ifcolonfoundonthisline\JSONstringvaluestyle\fi,
  MoreSelectCharTable =%
    \lst@DefSaveDef{`:}\colon@json{\processColon@json},
  basicstyle          = \ttfamily,
  keywordstyle        = \ttfamily\bfseries,
}

\newcommand\processColon@json{%
  \colon@json%
  \ifnum\lst@mode=\lst@Pmode%
    \global\colonfoundonthislinetrue%
  \fi
}

\lst@AddToHook{Output}{%
  \ifcolonfoundonthisline%
    \ifnum\lst@mode=\lst@Pmode%
      \def\lst@thestyle{\JSONnumbervaluestyle}%
    \fi
  \fi
  \lsthk@DetectKeywords%
}

\lst@AddToHook{EOL}%
  {\global\colonfoundonthislinefalse}
\makeatother

\newtcolorbox{myboxi}[1][]{
  breakable,
  title=#1,
  colback=red!5,
  colbacktitle=red!5,
  coltitle=black,
  fonttitle=\bfseries,
  bottomrule=0pt,
  toprule=0pt,
  leftrule=2pt,
  rightrule=2pt,
  titlerule=0pt,
  arc=0pt,
  outer arc=0pt,
  colframe=red,
}

\newtcolorbox{myboxnote}[1][]{
  breakable,
  title=#1,
  colback=orange!0,
  colbacktitle=orange!0,
  coltitle=black,
  fonttitle=\bfseries,
  bottomrule=0pt,
  toprule=0pt,
  leftrule=2pt,
  rightrule=2pt,
  titlerule=0pt,
  arc=0pt,
  outer arc=0pt,
  colframe=orange,
}

\definecolor{brightblue}{RGB}{33, 102, 172}
\newtcolorbox{promptbox}[1]{
    colback=white,   
    colframe=brightblue, 
    coltitle=white,         
    fonttitle=\bfseries,    
    width=0.84\textwidth,  
    center,               
    title={#1},            
    breakable
}

\newtcolorbox{myboxii}[1][]{
  breakable,
  freelance,
  title=#1,
  colback=white,
  colbacktitle=white,
  coltitle=black,
  fonttitle=\bfseries,
  bottomrule=0pt,
  boxrule=0pt,
  colframe=white,
  overlay unbroken and first={
  \draw[red!75!black,line width=3pt]
    ([xshift=5pt]frame.north west) --
    (frame.north west) --
    (frame.south west);
  \draw[red!75!black,line width=3pt]
    ([xshift=-5pt]frame.north east) --
    (frame.north east) --
    (frame.south east);
  },
  overlay unbroken app={
  \draw[red!75!black,line width=3pt,line cap=rect]
    (frame.south west) --
    ([xshift=5pt]frame.south west);
  \draw[red!75!black,line width=3pt,line cap=rect]
    (frame.south east) --
    ([xshift=-5pt]frame.south east);
  },
  overlay middle and last={
  \draw[red!75!black,line width=3pt]
    (frame.north west) --
    (frame.south west);
  \draw[red!75!black,line width=3pt]
    (frame.north east) --
    (frame.south east);
  },
  overlay last app={
  \draw[red!75!black,line width=3pt,line cap=rect]
    (frame.south west) --
    ([xshift=5pt]frame.south west);
  \draw[red!75!black,line width=3pt,line cap=rect]
    (frame.south east) --
    ([xshift=-5pt]frame.south east);
  },
}

\mdfdefinestyle{mystyle}{%
  rightline=true,
  innerleftmargin=10,
  innerrightmargin=10,
  outerlinewidth=3pt,
  topline=false,
  rightline=true,
  bottomline=false,
  skipabove=\topsep,
  skipbelow=\topsep
}

\usepackage{pifont}

\DeclareCaptionFont{black}{\color{black}}

\newenvironment{itemize*}%
 {\leftmargini=10pt\begin{itemize}%
  \setlength{\itemsep}{0pt}%
  \setlength{\parskip}{0pt}%
  }%
 {\end{itemize}}
\newenvironment{enumerate*}%
 {\begin{enumerate}%
  \setlength{\itemsep}{0pt}%
  \setlength{\parskip}{0pt}}%
 {\end{enumerate}}

\usepackage{etoolbox}
\newcounter{bibcount}
\makeatletter
\patchcmd{\@lbibitem}{\item[}{\item[\hfil\stepcounter{bibcount}{[\thebibcount]}}{}{}
\renewcommand\NAT@bibsetup%
  [1]{\setlength{\leftmargin}{\bibhang}\setlength{\itemindent}{-\parindent}%
      \setlength{\itemsep}{\bibsep}\setlength{\parsep}{\z@}}
\makeatother

\definecolor{myblue}{rgb}{0.9, 0.1, 0.94}
\definecolor{mygreen}{rgb}{0.64, 0.56, 0.88}
\definecolor{myyellow}{rgb}{0.68, 0.6, 0.1}
\definecolor{fancygreen}{rgb}{0.33, 0.68, 0.20}
\definecolor{salmon}{rgb}{0.94, 0.52, 0.49}
\definecolor{tablegreen}{rgb}{0.82, 0.94, 0.75}
\definecolor{tableblue}{rgb}{0.81, 0.90, 0.94}
\definecolor{tablered}{rgb}{0.97, 0.85, 0.85}
\definecolor{tableorange}{rgb}{0.96, 0.85, 0.81}

\begin{document}





\title{daVinci-Env: Open SWE Environment Synthesis at Scale}

\author[1,2,3]{Dayuan Fu\textsuperscript{*}}
\author[2,3]{Shenyu Wu\textsuperscript{*}}
\author[2,3]{Yunze Wu\textsuperscript{*}}
\author[2,3]{Zerui Peng\textsuperscript{*}}
\author[2,3]{Yaxing Huang\textsuperscript{*}}
\author[1]{Jie Sun\textsuperscript{*}}
\author[2,3]{Ji Zeng}
\author[1,2]{Mohan Jiang}
\author[2,3]{Lin Zhang}
\author[2]{Yukun Li} 
\author[2]{Minhao Liu}
\author[2]{Rui Ma}
\author[2]{Yueran Zang}
\author[2]{Zhaocheng Li}
\author[1]{Jiarui Hu}
\author[1]{Liming Liu}
\author[1]{Jinlong Hou\textsuperscript{†}}
\author[1,2,3]{Pengfei Liu\textsuperscript{†}}
\affil{SII \quad \textsuperscript{2}SJTU \quad \textsuperscript{3}GAIR}
\footnotetext[1]{* Equal contribution.}
\footnotetext[2]{† Corresponding authors.}

\maketitle

\pagestyle{fancy}
\thispagestyle{fancy}
\fancyhead{}
\lhead{
  \raisebox{-0.3cm}{\includegraphics[height=0.95cm]{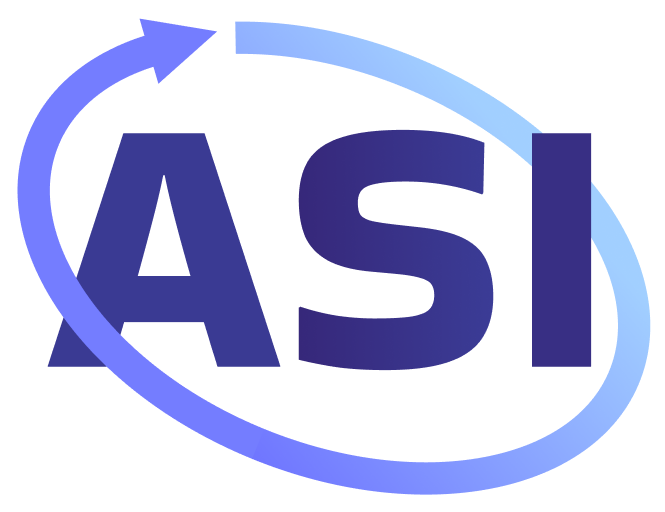}}
}
\rhead{%
  \raisebox{-0.2cm}{\includegraphics[height=0.7cm]{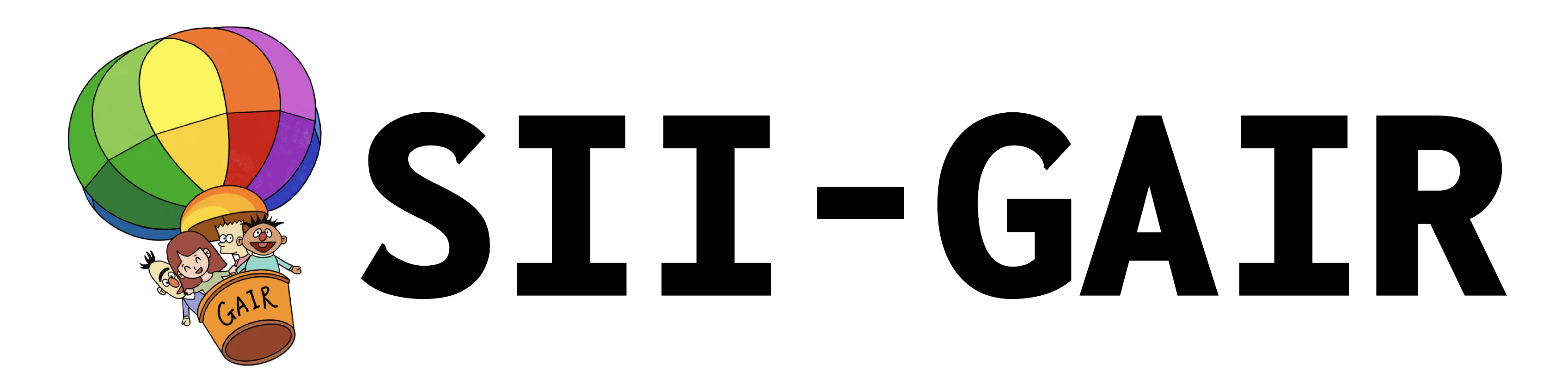}}%
}
\renewcommand{\headrulewidth}{0pt}
\setlength{\headsep}{2mm}

\begin{NoHyper}
\renewcommand{\thefootnote}{}
\end{NoHyper}
\vspace{-20pt}


\vspace{20pt}


\begin{abstract}

  Training capable software engineering (SWE) agents demands large-scale, executable, and verifiable environments that provide dynamic feedback loops for iterative code editing, test execution, and solution refinement. However, existing open-source datasets remain limited in scale and repository diversity, while industrial solutions are opaque with unreleased infrastructure, creating a prohibitive barrier for most academic research groups.
  We present \textbf{OpenSWE}, the largest fully transparent framework for SWE agent training in Python, comprising 45,320 executable Docker environments spanning over 12.8k repositories, with all Dockerfiles, evaluation scripts, and infrastructure fully open-sourced for reproducibility. OpenSWE is built through a multi-agent synthesis pipeline deployed across a 64-node distributed cluster, automating repository exploration, Dockerfile construction, evaluation script generation, and iterative test analysis. Beyond scale, we propose a quality-centric filtering pipeline that characterizes the inherent difficulty of each environment, filtering out instances that are either unsolvable or insufficiently challenging and retaining only those that maximize learning efficiency. With \$891K spent on environment construction and an additional \$576K on trajectory sampling and difficulty-aware curation, the entire project represents a total investment of approximately \$1.47 million, yielding about 13,000 curated trajectories from roughly 9,000 \textbf{quality guaranteed environments}.
  Extensive experiments validate OpenSWE's effectiveness: OpenSWE-32B and OpenSWE-72B achieve 62.4\% and 66.0\% on SWE-bench Verified, establishing SOTA among Qwen2.5 series. Models trained on OpenSWE consistently outperform those trained on SWE-rebench across all settings, with a log-linear data scaling trend showing no saturation. Moreover, SWE-focused training yields substantial \textbf{out-of-domain improvements}, including up to 12 points on mathematical reasoning and 5 points on science benchmarks, without degrading factual recall. All environments and evaluation scripts are publicly available at \url{https://github.com/GAIR-NLP/OpenSWE}.

\end{abstract}



\begin{figure}[h]
    \centering
    \includegraphics[width=0.93\linewidth]{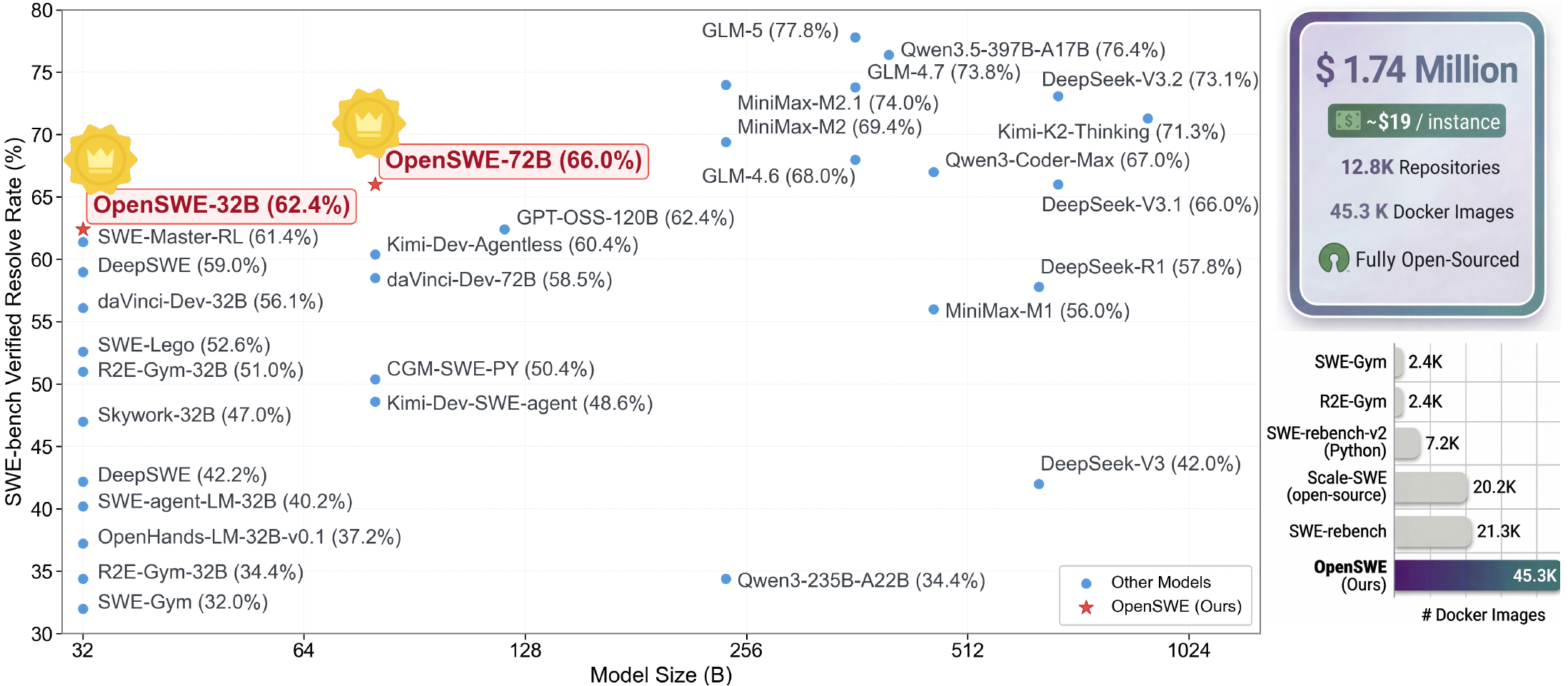}
    \caption{Image construction and performance overview of OpenSWE.}
    \label{fig:teaser}
\end{figure}

\newpage
\pagestyle{fancy}
\lhead{\rightmark}
\renewcommand{\headrulewidth}{0.7pt}
\setlength{\headsep}{5mm}
\clearpage

\newpage
\renewcommand{\thefootnote}{\arabic{footnote}}
\setcounter{footnote}{0}

\section{Introduction}
\label{sec:introduction}

The rapid advancement of Large Language Models (LLMs) has catalyzed the development of autonomous software engineering (SWE) agents~\citep{yang2024swe,team2025kimi,jiang2026davinci}. These systems can interpret complex requirements, navigate extensive codebases, iteratively edit code, run tests, and refine solutions without human intervention~\citep{fu2025agentrefine}. Unlike static code generation, these agents require \emph{verifiable and executable environments} like Docker~\citep{jimenez2023swe,xia2024agentless} to provide dynamic feedback loops: they must compile code, execute tests, and observe runtime behaviors to iteratively refine their solutions~\citep{yao2023react}.


However, constructing high-quality and diverse executable environments at scale remains a critical bottleneck. While recent open-source efforts such as SWE-rebench~\citep{badertdinov2025swerebenchautomatedpipelinetask}, SWE-Universe~\citep{chen2026swe}, and SWE-Factory~\citep{guo2026swefactoryautomatedfactoryissue} have made progress toward automation, the resource barrier is prohibitive: the computational and infrastructure costs of generating validated environments at scale remain extraordinarily high, effectively excluding most academic research groups and creating a stark divide between industrial solutions, which achieve scale but remain opaque with unreleased infrastructure~\citep{chen2026swe,liu2025deepseek}, and open-source alternatives that remain limited in both scale and repository diversity.

\begin{figure}[t]
    \centering
    \includegraphics[width=1.\linewidth]{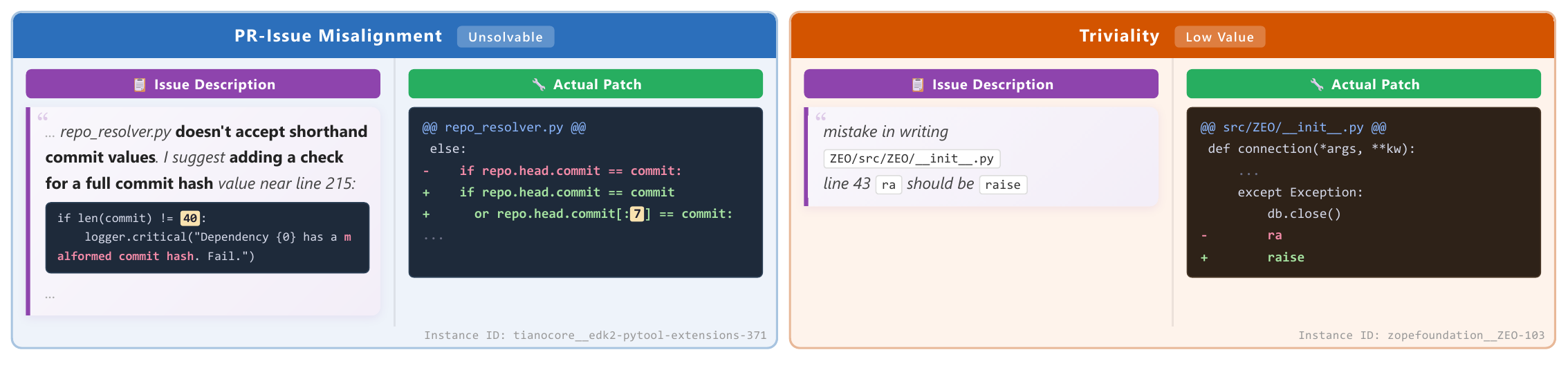}
    \caption{Two specific risks in SWE tasks. \textbf{Left}: The PR is unsolvable because the first seven characters of the commit hash can pass the test, whereas the issue requires checking the full hash. \textbf{Right}: The PR is trivial, since the issue have just tell the modified file and the string that should be changed.}
    \label{fig:badcase}
\end{figure}

Beyond the cost of environment construction, the quality and difficulty distribution of these environments are equally critical for effective agent training. While scaling the number of environments is a necessary condition, it is far from sufficient on its own. As illustrated in Figure~\ref{fig:badcase}, environments synthesized from real repositories frequently suffer from PR-Issue misalignment, where the submitted patch does not actually resolve the described issue, or triviality, where the issue description directly reveals the solution. Such environments are either effectively unsolvable or too simple to provide meaningful learning signal. More broadly, the difficulty distribution across environments plays a decisive role in training effectiveness, and identifying the subset at appropriate difficulty levels that maximizes learning efficiency requires systematic evaluation and careful curation.

In this work, we address both challenges by introducing \textbf{OpenSWE}, the largest fully transparent framework for SWE agent training to date. OpenSWE comprises 45,320 executable Docker environments spanning 12.8k repositories, representing over \$891,000 in construction costs, \emph{with all Dockerfiles, evaluation scripts, and distributed infrastructure fully open-sourced}. Unlike prior work, we release not only the final environments but also the complete synthesis pipeline: a multi-agent system deployed across a 64-node cluster that automates repository exploration, Dockerfile construction, evaluation script generation, and iterative test analysis. To ensure data quality beyond mere scale, we propose a quality-centric filtering pipeline that characterizes the inherent difficulty of each environment, filtering out those that are either unsolvable or insufficiently challenging and retaining only environments at appropriate difficulty levels that provide the most effective learning signal. This large-scale trajectory sampling and curation process requires an additional computational investment of approximately \$576,000, ultimately yielding about 13,000 curated trajectories from a subset of roughly 9,000 high-quality environments.

Extensive experiments on these trajectories validate the effectiveness of OpenSWE and highlight the complementary roles of data scaling and difficulty-aware curation. Models trained on our curated trajectories achieve 62.4\% (32B) and 66.0\% (72B) on SWE-Bench Verified, establishing state-of-the-art among supervised fine-tuning methods and consistently outperforming SWE-rebench-trained models across all configurations. Data scaling analysis reveals a log-linear improvement trend with no saturation, confirming that additional high-quality environments continue to yield meaningful gains. Equally important, difficulty-aware filtering contributes measurably beyond raw scale: by retaining environments at the appropriate difficulty frontier, training efficiency improves significantly compared to using all environments indiscriminately. Furthermore, training on OpenSWE yields substantial out-of-domain improvements, including up to 12 points on mathematical reasoning and up to 5 points on science benchmarks, without degrading factual recall.

The specific contributions of this work are:
\begin{itemize*}
\item \textbf{Unprecedented Scale with Full Transparency:} We release 45,320 executable environments from 12.8k repositories at a construction cost of \$891K, with complete infrastructure including all Dockerfiles, evaluation scripts, and the distributed synthesis pipeline, enabling reproducibility and community-driven improvements.
\item \textbf{Quality-Centric Filtering via Difficulty-Aware Curation:} We propose a filtering pipeline that characterizes environment difficulty to filter out unsolvable and trivially simple instances. With an additional \$576K investment in trajectory sampling and curation, we obtain about 13,000 curated trajectories from roughly 9,000 high-quality environments.
\item \textbf{Strong Empirical Validation with Scaling and Curation Insights:} OpenSWE-trained models establish new SOTA results (62.4\%/66.0\%) among SFT methods under Qwen2.5 series, consistently outperform SWE-rebench across all scales and scaffolds, and exhibit log-linear scaling with no saturation. Both data scaling and difficulty-aware filtering are shown to be essential and complementary drivers of agent performance.
\end{itemize*}

\section{Related Work}

\subsection{Environment Synthesis} The construction of executable environments for agents has become a central infrastructure challenge. SWE-bench~\citep{jimenez2023swe} pioneered this direction by curating a benchmark of real GitHub issues paired with pull requests, where each task instance is embedded in a Docker-based repository snapshot with executable test suites that serve as evaluation oracles. To overcome this bottleneck, several concurrent efforts have emerged to automate large-scale environment generation. SWE-rebench~\citep{badertdinov2025swerebenchautomatedpipelinetask} introduces a scalable pipeline that replicates the SWE-bench construction process across a broader set of repositories, aiming to generate thousands of additional task instances with executable test environments. SWE-Universe~\citep{chen2026swe} takes a complementary approach by systematically crawling and filtering GitHub repositories to produce a diverse universe of candidate environments. SWE-Factory~\citep{guo2026swefactoryautomatedfactoryissue} and  Scale-SWE~\citep{zhao2026immersiongithubuniversescaling} further automate the end-to-end pipeline from repository selection to Dockerfile synthesis and test harness generation. Scale-SWE further scales this paradigm through a sandboxed multi-agent workflow. BeyondSWE~\citep{chen2026beyondswecurrentcodeagent} expands the evaluation scope beyond single-repository bug fixing by introducing more complex real-world scenarios such as cross-repository reasoning, dependency migration, and domain-specific development tasks.SWE-World~\citep{sun2026swe} proposes an orthogonal direction by replacing physical Docker execution with learned surrogate models trained on agent-environment interaction data, eliminating the resource-intensive costs of Docker environment maintenance while preserving the agent-environment feedback loop.

\subsection{SWE Agents Training} The development of autonomous software engineering agents has progressed rapidly from simple code completion to complex, multi-step task resolution in real-world repositories. To enable LLMs to interact effectively with these repositories, agent scaffolds have emerged as critical infrastructure. SWE-agent~\citep{yang2024swe} serves as a foundational example, establishing a baseline where agents can autonomously navigate codebases, localize bugs, and generate patches. Building on similar architectural principles, OpenHands~\citep{wang2025openhands} provides an extensible open-source platform utilizing the CodeAct framework, which allows agents to interleave code execution and natural language reasoning within a unified action space.

On the training and data synthesis side, SWE-smith~\citep{yang2025swesmithscalingdatasoftware} constructs a large-scale training data synthesis pipeline that generates diverse task instances and execution trajectories for supervised fine-tuning of SWE agents, enabling the training of open-weight SWE agents from scratch. daVinci-Dev~\citep{zeng2026davinci} takes a different approach by combining structured planning with iterative code generation and debugging, leveraging multi-step reasoning traces to produce high-quality resolution trajectories.  SWE-Fixer~\citep{xie2025swe} focuses on scaling supervised fine-tuning with filtered, high-quality resolution trajectories. The SWE-Master~\citep{song2026swe} technical report systematically compares these representative approaches.


\section{Method}


\subsection{Github PR Collection}
We collect GitHub PRs from a broad set of Python repositories through GitHub REST\footnote{\url{https://docs.github.com/en/rest}} and GraphQL APIs  .\footnote{\url{https://docs.github.com/en/graphql}}
For each repository, we obtain PR metadata and selectively query additional endpoints for detailed content, including linked issue descriptions when available, and the full commit sequence with corresponding diffs.

\begin{figure}[t]
    \centering
    \includegraphics[width=0.95\linewidth]{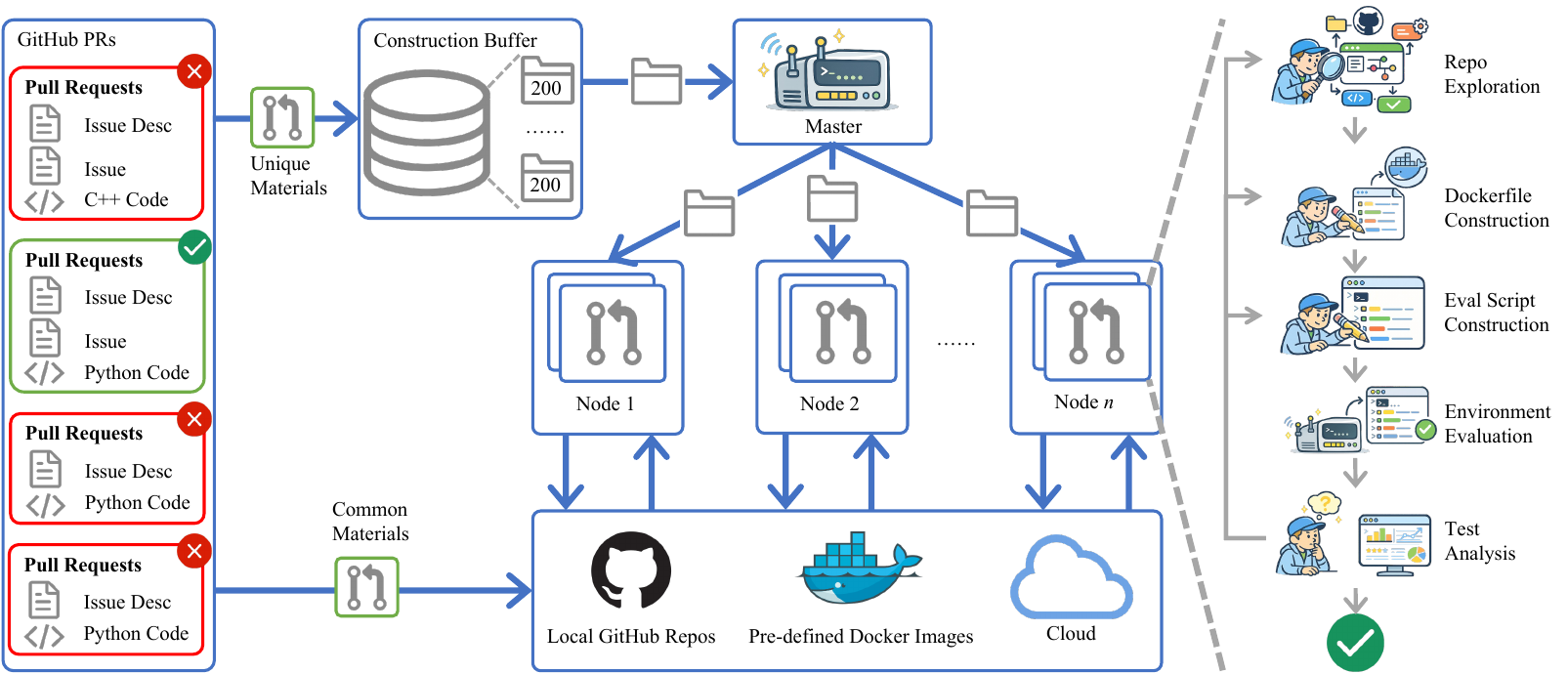}
    \caption{The framework of OpenSWE.}
    \label{fig:framework}
\end{figure}

\subsection{GitHub PR Filtering}\label{github-pr-filtering}

The filtering process operates on the GitHub PR dataset obtained through the collection pipeline described above. Each entry comprises four essential fields: repository identifier, PR number, associated issues, and the complete PR patch encompassing all code modifications.

To guarantee the quality and suitability of PRs, we apply a four-stage filtering pipeline:

\paragraph{Repository Viability.} To improve the representativeness of our dataset, we retain only repositories with at least five GitHub stars, using star count as a proxy for community validation and project maturity. This criterion excludes nascent or unmaintained projects that are unlikely to reflect real-world software engineering practice.

\paragraph{Language Filter.} We constrain the dataset to PRs from repositories whose primary programming language is Python, as determined by GitHub's language detection. This aligns with the predominant language coverage in existing code generation benchmarks and ensures evaluation consistency.

\paragraph{Issue Requirement.} Since every task should be grounded in a well-defined natural language problem statement, each PR is required to have at least one associated issue with an issue description. PRs lacking linked issues or containing only empty issue descriptions are excluded due to the absence of sufficient task specification.

\paragraph{Substantive Code Changes.} In order to guarantee that each instance tests real implementation ability rather than auxiliary testing effort, we require non-empty patches to non-test code and exclude PRs whose changes are confined entirely to test directories or test files (like \texttt{*tests*}, \texttt{*spec*}, or \texttt{*e2e*} in its path).

After identifying high-quality PR candidates, we use a multi-agent system to transform the selected PRs into real SWE environments. Each environment requires a reproducible Docker container with the correct dependencies, as well as a validated evaluation script capable of confirming whether an agent's solution is correct.

\subsection{Repository Exploration}
\label{repo-exploration}

We introduce a lightweight repository exploration agent that bridges raw repository state and downstream environment generation. The agent is initialized with repository-level metadata (repository name, commit/version, and patch-derived file cues) and performs bounded exploration over the local checkout to collect only setup- and test-relevant evidence for subsequent agents.

\paragraph{Targeted Retrieval Interface.}
The agent operates through three constrained repository APIs: (1) \texttt{browse} for structural inspection, (2) \texttt{search} for locating candidate configuration files, and (3) \texttt{digest} for extracting actionable setup and test instructions from selected files. This interface is intentionally narrow, encouraging low-cost retrieval centered on high-yield artifacts such as \texttt{README.md}, \texttt{CONTRIBUTING.md}, dependency manifests, and CI workflows.

\paragraph{Cost-Aware Iterative Policy.}
Exploration proceeds in multiple rounds and follows a conservative policy: in the absence of explicit failure feedback, the agent performs shallow, document-first inspection; when the test analysis agent reports missing context, retrieval is redirected to only the requested files or configuration dimensions. This design reduces redundant repository traversal while preserving the ability to recover from environment or test-command ambiguity in later iterations.

\paragraph{Minor Implementation Details.}
We include several small implementation details in this stage: (1) the extraction scope explicitly captures Python-specific environment-management frameworks (e.g., \texttt{poetry}, \texttt{uv}) in addition to test frameworks, to help the Docker construction agent retrieve enough context in advance; and (2) API-call parsing and argument validation are enclosed in exception-safe handling to prevent malformed invocations from terminating retrieval rounds.

\subsection{Dockerfile Construction}\label{dockerfile-construction}

The Dockerfile agent is responsible for generating an environment for each task. 
During the pilot study, we identified two recurring failure modes: (1) network instability during environment construction, where generic base images require downloading Python and dependencies at build time, leading to frequent timeouts; and (2) redundant rebuilds, where unchanged base layers are reconstructed from scratch on every iteration. 
These inefficiencies become particularly costly at scale; therefore, we equip the Dockerfile agent with the following strategies.

\paragraph{Base Image Strategy.} Rather than starting from generic Ubuntu images that require runtime Python installation, we pre-build a suite of \texttt{openswe-python} base images covering Python 2.7 and 3.5--3.14, each bundled with a conda package, a pre-activated \texttt{testbed} environment, and configured package mirrors for reliability. This eliminates the most common source of build failures---network timeouts during dependency installation---and enables immediate layer reuse across tasks sharing the same Python version.

\paragraph{Repository Provisioning.} Instead of cloning repositories inside the container at build time, we maintain a local bare repository cache and inject the codebase via \texttt{COPY}, with each task's target commit checked out in advance. This removes GitHub API rate limits and network failures from the agent loop entirely and improves reproducibility by eliminating dependence on external availability. It also reduces the error rate of the agent by avoiding the repetition of long commit hashes.

\paragraph{Layer-Aware Prompting and Python-Specific Optimizations.} We observe that in typical agentic workflows, dependency specifications are revised far more frequently than the Dockerfile structure itself. Leveraging this observation, we explicitly instruct the agent to place stable base layers early in the Dockerfile so they are cached by Docker, and to isolate dependency installation into later layers that can be cheaply rebuilt across iterations. This yields significant speedups when the agent iterates on dependency fixes without altering the base environment. Prompts also enforce Python-specific correctness requirements, including proper conda environment activation, development-mode package installation, and deferred test execution to the evaluation script.

The Dockerfile agent will receive the repository exploration agent's findings (e.g., special dependencies from \texttt{README.md}) as additional input, allowing the agent to make more informed initial decisions, and it will operate iteratively to construct the Dockerfile. If the final test execution fails, the Dockerfile agent will also receive the feedback from the test analysis agent and refine its output in subsequent attempts.

\subsection{Evaluation Script Construction}\label{evaluation-script-construction}

The evaluation script agent generates bash scripts that verify repair correctness by executing tests and confirming that failures introduced by the issue can be resolved by the patch under evaluation. The central challenge is \textbf{precise test targeting}: only the test cases directly relevant to the issue should be executed. Accordingly, the agent identifies the specific test files tied to the issue and, when necessary, synthesizes new test cases to cover scenarios not present in the original PR.

\paragraph{Test Design.} Because the agent may introduce new test cases beyond those in the original PR, the static fail2pass scripts used in SWE-Bench are no longer applicable. We instead instruct the agent to construct a structured bash script from scratch, incorporating: (1) the selected and synthesized test cases with correct exit code capture; (2) output delimiters marking the start and end of test output for reliable log parsing; and (3) a dedicated exit code marker ($\texttt{OPENSWE\_EXIT\_CODE}$) embedded in the script output, whose value serves as the final signal for determining repair correctness.

\paragraph{Script Design.} To support stable iteration, the script is template-based, separating patch injection from test command logic so that the agent can refine test invocations across iterations without regenerating the entire script. For conda-based environments, explicit activation sequences are enforced to prevent subtle \texttt{PATH} issues that would silently corrupt test results.

Like the Dockerfile agent, the evaluation script agent operates within the same iterative feedback loop: the repository exploration agent and Dockerfile agent supply repository context prior to generation, and after test execution, the test analysis agent inspects the final result of the test execution and determines whether the repair is correct. If not, it will provide feedback to the evaluation script agent to refine the script for the next iteration.

\subsection{Environment Evaluation}\label{environment-evaluation}

With the Dockerfile and evaluation script in place, the pipeline proceeds to rule-based validation. For each iteration, the Docker image is built once and the evaluation script is executed under two conditions: first applying a test-only patch to verify that the tests indeed fail on the unpatched codebase, then applying the full fix patch to verify that all tests pass. A sample is accepted only when both conditions are met. The exit code marker \texttt{OPENSWE\_EXIT\_CODE=X} is parsed from script output via regex; if the marker is absent, validation is marked as failed and targeted feedback is returned to the agent.

To support this validation at scale, we introduce two infrastructure optimizations. First, to ensure reproducible results and prevent resource contention across concurrent evaluations, each container is bound to 4 dedicated CPU cores, a 24\,GB memory cap, and a 200\,GB storage limit. Second, rather than discarding images after every iteration, we retain images until the Dockerfile changes---yielding a 5$\times$ speedup in the common case where only the evaluation script is revised. Successfully validated images are pushed to a remote registry for reuse in subsequent training and evaluation.

\subsection{Test Analysis}\label{test-analysis}

Once the rule-based validation completes, the test analysis agent examines the results regardless of whether the sample passed or failed. On passing results, it inspects the logs to verify that the success is genuine---checking that the evaluation script does not contain hardcoded exit codes or other shortcuts that bypass real test execution. On failures, it diagnoses the root cause: a Dockerfile misconfiguration, an evaluation script error, or an inherently unsolvable environment (e.g., conflicting dependencies, unavailable Python versions). For fixable errors, it generates targeted feedback that is routed back to the responsible agent for the next iteration; for inherently unsolvable cases, it marks the sample to enable early exit. The final dataset retains only samples that pass both the rule-based evaluation and the agent's legitimacy check.

\subsection{Multi-Machine Construction}

To facilitate the large-scale synthesis described in Section~\ref{sec:introduction}, we deployed a distributed computing cluster comprising 64 Elastic Compute Service (ECS) instances. This infrastructure enables the simultaneous processing of an extensive corpus of approximately 572,114 GitHub PRs by parallelizing the Docker-based evaluation pipeline across isolated nodes. We use Deepseek-v3.2~\cite{liu2025deepseek} as the construction model.

Constructing environments at this scale presents significant engineering challenges: 
\begin{itemize*}
    \item \textbf{Execution Instability}: The pipeline relies on non-deterministic external factors, including LLM API latency, network-dependent dependency resolution, and the execution of agent-synthesized scripts, all of which can lead to unexpected process crashes.
    \item \textbf{Resource Contention}: Standard Docker engines lack the granular resource isolation required to prevent memory exhaustion (OOM) or disk saturation during intensive builds, potentially destabilizing the host node.
\end{itemize*}

To address these, we designed a decoupled, fault-tolerant parallelization framework:

\begin{itemize*}
    \item \textbf{Data Parallelism with Minimal Coupling}: We adopted a data-parallel approach to minimize inter-node dependencies. Unlike tightly coupled frameworks such as MPI or Ray, where a single node failure can halt the entire job, our architecture ensures that nodes operate independently.
    \item \textbf{Shared Filesystem Message Queue}: Communication and task distribution are managed through a file-based message queue hosted on a shared filesystem. This design decouples the task producer from the consumers, ensuring that individual node failures do not result in data loss or system-wide paralysis.
    \item \textbf{Resilient Process Management}: All synthesis processes are managed via \texttt{systemd} services. This configuration provides automated service recovery and restarts in the event of unexpected software termination.
    \item \textbf{Automated Resource Pruning}: To prevent storage and memory exhaustion from ``zombie" containers or orphaned images—frequent side effects of interrupted agent scripts—we implemented an automated cleanup daemon that aggressively prunes unused Docker resources.
    \item \textbf{Observability and Monitoring}: We deployed a monitoring stack based on Prometheus and Grafana to track performance metrics and task progress in real-time, allowing for rapid diagnosis of hardware or pipeline anomalies.
\end{itemize*}

The hardware and software specifications for each of the 64 compute nodes are standardized in Table~\ref{tab:hardware_config}. Through empirical experiments in a small scale, we identified this per-node specification as a near-optimal operating point: it provides sufficient per-task throughput while avoiding the diminishing returns observed with further resource scaling. With this 64-node cluster, we completed the construction of 45,320 validated environments in approximately two weeks, reducing what would otherwise be a months-long process and making iterative refinement of the synthesis pipeline practically feasible.


\subsection{Environment Statistics}

Table~\ref{tab:env_stats} compares OpenSWE against existing SWE training datasets in terms of scale and executability. We filtered all instances that have been created in SWE-rebench and SWE-Bench Verified.
OpenSWE provides the largest number of executable repositories and tasks among all datasets, covering 12.8k repos and 45.3k tasks.

\begin{table}[t]
\centering
\begin{tabular}{ll}
\hline
\textbf{Component} & \textbf{Specification} \\ \hline
\multicolumn{2}{l}{\textit{Hardware Configuration}} \\
CPU & Intel(R) Xeon(R) 6982P-C (32 Virtualized Cores) \\
Memory & 128 GB RAM \\
Network & 20 Gbps Intranet Bandwidth \\
Storage & 4 TB SSD \\ \hline
\multicolumn{2}{l}{\textit{Software Configuration}} \\
Operating System & Ubuntu 24.04 LTS \\
Container Engine & Docker 29.1.3 \\ \hline
\end{tabular}

\caption{Hardware and Software Specifications for Distributed Synthesis Nodes.}
\label{tab:hardware_config}
\end{table}

\begin{table}[t]
\centering
\resizebox{0.8\linewidth}{!}{
\begin{tabular}{l c c c c}
\toprule
\textbf{Dataset} & \textbf{\# Repos} & \textbf{\# images} & \textbf{\# Tasks}  & \textbf{Source} \\
\midrule
R2E-Gym (Subset) \citep{jain2025r2egymproceduralenvironmentshybrid}        & 10   & 2.4k & 4.6k    & Synthetic \\
SWE-gym \citep{pan2024training}                  & 11   & 2.4k  & 2.4k   & Real \\
SWE-rebench  \citep{badertdinov2025swerebenchautomatedpipelinetask}                                & 3.5k  & 21.3k  & 21.3k         & Real \\
SWE-rebench (filtered)                         & 3.3k  & 18.8k & 18.8k & Real \\
SWE-rebench-v2  \citep{badertdinovSWErebenchV2LanguageAgnostic2026}                                & 2.7k  & 32.7k  & 32.7k         & Real \\
SWE-rebench-v2 (Python)                           & 573  & 7.2k  & 7.2k         & Real \\
Scale-SWE~\citep{zhao2026immersiongithubuniversescaling}                                & 5.2k  & 100k  & 100k        & Real \\
Scale-SWE (open-sourced)                           & 1.2k  & 20.2k  & 20.2k         & Real \\
\midrule
\textbf{OpenSWE (ours)}       & 12.8k & 45.3k & 45.3k  & Real \\
\bottomrule
\end{tabular}
}

\caption{Comparison of SWE training environment.  SWE-rebench is filtered because some environments fail to execute the gold patch under our infrastructure.}
\label{tab:env_stats}
\end{table}

\subsection{Training}

\paragraph{Training Data Collection} To construct our training data, we used the GLM-4.7 model to sample the trajectory from the entire OpenSWE and SWE-rebench (filtered) datasets four times under the OpenHands or SWE-Agent (temperature 1.0, 200k context, and 300 steps). We then collected all trajectories that were correct in one or two of the four attempts under the same instance. To ensure training quality, we masked any steps that contained formatting errors or other mistakes, leading to an error observation. We also remove all data that contains 'git pull' in the bash action to avoid reward hacking. 

\paragraph{SFT Training}  We modified the slime code\footnote{\url{https://github.com/THUDM/slime}} to support multiturn training with correct action masking. All models are trained with a max token of 128k, 5 epochs, batch size 128, and a learning rate from 1e-5 to 1e-6 with cosine annealing. We use Qwen2.5-32B-Base and Qwen2.5-72B-Base as our base models.

\section{Experiments}

\subsection{Experimental Setup}
We evaluate our model on SWE-Bench Verified using OpenHands or SWE-Agent (temperature 0.7, 128k context, and 300 steps) and report Pass@1, averaged across 2 runs.

\begin{table}[t]
\centering
\resizebox{0.90\linewidth}{!}{
\begin{tabular}{l c c c}
\toprule
\textbf{Model} & \makecell{Backbone} & \textbf{Scaffold} & \textbf{Score} \\
\midrule
\multicolumn{4}{l}{\textit{Qwen 2.5 32B Coder Series}} \\
R2EGym-Agent~\citep{jain2025r2egymproceduralenvironmentshybrid} & Qwen2.5-32B-Coder-Base  & R2E-Gym & 34.4 \\
Openhands-LM~\citep{wang2025openhands} & Qwen2.5-Coder-32B-Inst.   & OpenHands & 37.2 \\
SWE-Agent-LM~\citep{yang2025swesmithscalingdatasoftware} & Qwen2.5-Coder-32B-Inst. & SWE-Agent & 40.2 \\
SWE-Mirror-LM~\citep{wang2025swemirrorscalingissueresolvingdatasets} & Qwen2.5-Coder-32B-Inst. & MOpenHands & 52.2 \\
Skywork-SWE~\citep{zeng2025skyworksweunveilingdatascaling} & Qwen2.5-Coder-32B-Inst.  & OpenHands & 38.0 \\
SWE-Compressor~\citep{liu2025contexttoolcontextmanagement} & Qwen2.5-32B-Base  & OpenHands & 57.6 \\
SWE-Master-32B~\citep{song2026swe} &  Qwen2.5-Coder-32B-Inst. &  R2E-Gym &   57.8 \\
SWE-Master-32B-RL~\citep{song2026swe} &  Qwen2.5-Coder-32B-Inst. &  R2E-Gym &  61.4 \\
\midrule
\multicolumn{4}{l}{\textit{Qwen 3 30B-A3B Series}} \\
Qwen3-30B-A3B-Instruct~\citep{zhao2026immersiongithubuniversescaling}  & Qwen3-30B-A3B-Instruct  & OpenHands & 22.0 \\
Scale-SWE~\citep{zhao2026immersiongithubuniversescaling}  & Qwen3-30B-A3B-Instruct  & OpenHands & 64.0 \\
\midrule
\multicolumn{4}{l}{\textit{Qwen 3 32B Series}} \\
FrogBoss~\citep{sonwane2025bugpilotcomplexbuggeneration}  & Qwen3-32B  & SWE Agent & 54.6 \\
SWE-Lego-Qwen3-32B~\citep{tao2026swelegopushinglimitssupervised}  & Qwen3-32B  & OpenHands & 52.6 \\
CoderForge-32B~\cite{CoderForge2026} & Qwen3-32B  & OpenHands & 	59.4 \\
\midrule
\multicolumn{4}{l}{\textit{Qwen 2.5 32B Series}} \\

daVinci-Dev-32B~\cite{zeng2026davinci} & Qwen2.5-32B-Base  & SWE-Agent & 56.1 \\
OpenSWE-32B (Ours)& Qwen2.5-32B-Base  & OpenHands & 59.8 \\
OpenSWE-32B (Ours)& Qwen2.5-32B-Base  & SWE-Agent & \textbf{62.4} \\
\midrule
\multicolumn{4}{l}{\textit{Qwen 2.5 72B Series}} \\
SWE-Fixer-72B~\cite{xie2025swe} & Qwen2.5-72B-Base & Agentless & 32.8 \\
daVinci-Dev-72B~\cite{zeng2026davinci} & Qwen2.5-72B-Base & SWE-Agent & 58.5 \\
Kimi-Dev~\citep{yang2025kimidevagentlesstrainingskill} & Qwen2.5-72B-Base & Agentless & 60.6 \\
OpenSWE-72B (Ours)& Qwen2.5-72B-Base & OpenHands & 65.0 \\
OpenSWE-72B (Ours)& Qwen2.5-72B-Base & SWE-Agent & \textbf{66.0} \\
\bottomrule
\end{tabular}
}
\caption{Comparison with representative methods on \texttt{SWE-Bench Verified}. We include representative works with agentic scaffolds.}
\label{tab:main_comparison}
\end{table}

\subsection{Main Results}

Table~\ref{tab:main_comparison} presents the comparison of OpenSWE with representative on SWE-Bench Verified. 

\paragraph{State-of-the-Art at Both Scales} OpenSWE-32B achieves a resolution rate of 62.4\%, surpassing all methods on Qwen2.5 series. Compared to the strongest Qwen2.5-Coder-32B baseline SWE-Master-32B and SWE-Master-32B-RL. OpenSWE-32B improves by 4.6\% while using a non-Coder base model, demonstrating that high-quality environment data can compensate for domain-specific pretraining. At the 72B scale, OpenSWE-72B reaches 66.0\%, outperforming daVinci-Dev-72B by 7.5\%. Both the 32B result and 72B result prove the effectiveness of OpenSWE.

\paragraph{Scaling with Model Capacity} OpenSWE-72B improves over OpenSWE-32B by 3.6\%. In contrast, for daVinci-Dev, scaling from 32B to 72B yields only a 2.4\% gain with the same scaffold, suggesting that higher-quality training environments enable models to better leverage increased parameters. 

\paragraph{Scaffold-Agnostic Effectiveness}  As shown in Table~\ref{tab:main_comparison}, OpenSWE-32B reaches 59.8\% with OpenHands and 62.4\% with SWE-Agent; OpenSWE-72B reaches 65.0\% with OpenHands and 66.0\% with SWE-Agent. This indicates that high-quality environment data benefits multiple scaffold designs rather than being tied to a specific agent framework, enhancing the practical applicability of our approach.

\subsection{Data Scaling Analysis}

\begin{figure}[t]
    \centering
    \includegraphics[width=0.95\linewidth]{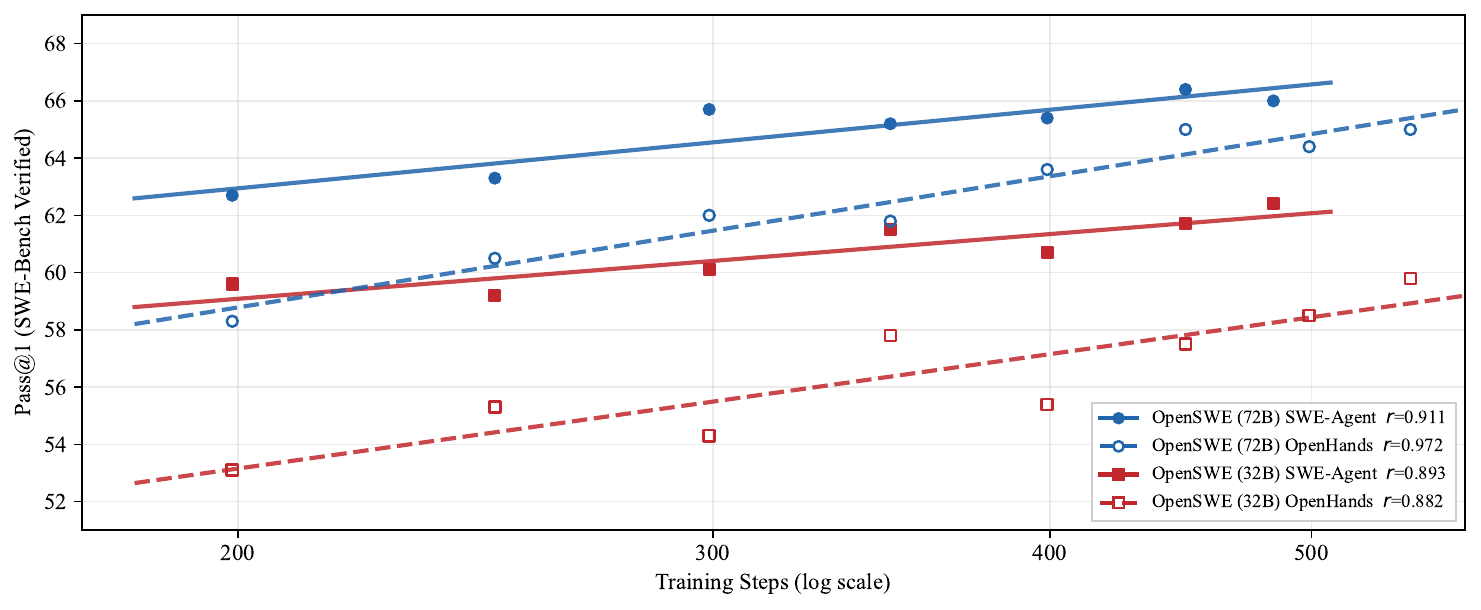}
    \caption{Data scaling curves for OpenSWE across model sizes and agent scaffolds in log-linear mode. Filled markers with solid fit lines denote SWE-Agent; hollow markers with dashed fit lines denote OpenHands. Blue indicates 72B models; red indicates 32B models.}
    \label{fig:data_scaling}
\end{figure}

To investigate the effect of training data scale on agent performance, we construct subsets of varying sizes from the full OpenSWE training set and evaluate checkpoints across two model scales (Qwen2.5-32B and Qwen2.5-72B) and two agent scaffolds (SWE-Agent and OpenHands). Figure~\ref{fig:data_scaling} presents the results.

\paragraph{Log-Linear Scaling Trend} Across all four model--scaffold configurations, Pass@1 improves approximately log-linearly with training steps. We fit a linear model in log-step space for each curve and observe consistently high Pearson correlation coefficients: $r{=}0.972$ for 72B CodeAct, $r{=}0.911$ for 72B SWE-Agent, $r{=}0.893$ for 32B SWE-Agent, and $r{=}0.882$ for 32B CodeAct. The uniformly high $r$ values across both model sizes and both scaffolds suggest that the log-linear scaling behavior is a robust property of the training data rather than an artifact of a specific architecture or evaluation protocol.

\paragraph{Larger Models Benefit More from Scaling} The 72B models consistently outperform their 32B counterparts across all training steps. Moreover, the gap widens as training progresses: at early checkpoints, the 72B SWE-Agent leads the 32B SWE-Agent by approximately 3.1\%, while at $\sim$484 steps this gap grows to 3.6\%. More notably, for the CodeAct scaffold, the 72B model improves from a 5.2\% lead at step 199 to a 5.2\% lead at step 544, indicating that larger models extract greater benefit from additional training data.

\paragraph{Scaffold Comparison} SWE-Agent consistently outperforms OpenHands across both model scales. For the 72B model, SWE-Agent achieves 66.0\% at the final checkpoint compared to OpenHands's 65.0\%; for the 32B model, SWE-Agent reaches 62.4\% versus OpenHands's 59.8\%. This 1--3\% margin suggests that the SWE-Agent scaffold's design provides a consistent advantage, though both scaffolds benefit similarly from data scaling.

\paragraph{No Saturation Observed} Importantly, none of the four curves show signs of saturation within our current budget. The continued upward trend at the largest training step counts suggests that further scaling the OpenSWE training set would yield additional performance gains, motivating future work on even larger-scale environment synthesis.

\subsection{Impact of Environment Source}

To understand how the choice of environment affects downstream agent performance, we train identical models on environments from different sources and evaluate under the same protocol. Table~\ref{tab:env_source} reports the results.

\begin{table}[t]
\centering

\begin{tabular}{l cc cc}
\toprule
 & \multicolumn{2}{c}{\textbf{SWE-Agent}} & \multicolumn{2}{c}{\textbf{CodeAct}} \\
\cmidrule(lr){2-3} \cmidrule(lr){4-5}
\textbf{Training Data} & 32B & 72B & 32B & 72B \\
\midrule
SWE-rebench        & 50.2\% & 63.4\% & 51.4\% & 62.4\% \\
OpenSWE             & 62.4\% & 66.0\% & 59.8\% & 65.0\% \\
SWE-rebench + OpenSWE   & 61.4\% & 68.0\% & 60.3\% & 65.5\% \\
\bottomrule
\end{tabular}

\caption{Impact of environment source on SWE-Bench Verified Pass@1 (\%) across model sizes and scaffolds.}
\label{tab:env_source}
\end{table}

\paragraph{OpenSWE Environments Are Substantially More Effective} Training on OpenSWE alone yields large improvements over SWE-Rebench across all four configurations. The most pronounced gain appears at the 32B SWE-Agent setting, where OpenSWE outperforms SWE-Rebench by 12.2\% absolute (62.4\% vs.\ 50.2\%). Even for the 72B CodeAct configuration where SWE-Rebench is most competitive, OpenSWE still leads by 2.6\% (65.0\% vs.\ 62.4\%). This demonstrates that the quality and diversity of OpenSWE's synthesized environments provide a stronger training signal than SWE-Rebench.

\paragraph{Complementary Value of Mixing Sources} Combining SWE-Rebench with OpenSWE yields further gains for 72B models: the 72B SWE-Agent configuration reaches 68.0\%, a 2.0\% improvement over OpenSWE alone and the best result across all settings. This suggests that SWE-Rebench introduces complementary environment patterns that benefit larger models. However, for 32B models, mixing slightly degrades performance on SWE-Agent (61.4\% vs.\ 62.4\%), indicating that smaller models may be more sensitive to distribution shifts introduced by heterogeneous data sources.

\paragraph{Robustness Across Scaffolds} The relative ordering of environment sources is consistent across both SWE-Agent and CodeAct scaffolds: OpenSWE consistently outperforms SWE-Rebench, and mixing provides additional gains primarily for larger models. This scaffold-agnostic pattern reinforces that the performance differences stem from the quality of the training environments rather than scaffold-specific interactions.

\subsection{General Capability Evaluation}
\label{sec:general_capability}

\begin{table}[h]
\centering
\small
\setlength{\tabcolsep}{4pt}

\begin{tabular}{lcccccc}
\toprule
 & \multicolumn{3}{c}{\textbf{Qwen2.5-32B}} & \multicolumn{3}{c}{\textbf{Qwen2.5-72B}} \\
\cmidrule(lr){2-4} \cmidrule(lr){5-7}
\textbf{Benchmark} & \textbf{Base} & \textbf{OpenSWE} & $\Delta$ & \textbf{Base} & \textbf{OpenSWE} & $\Delta$ \\
\midrule
\multicolumn{7}{l}{\textit{Code Benchmarks}} \\
HumanEval~\citep{chen2021codex}      & 61.43 & 90.52 & \textcolor{orange}{+29.09} & 66.82 & 76.25 & \textcolor{orange}{+9.43} \\
HumanEval+~\citep{evalplus}     & 54.01 & 85.24 & \textcolor{orange}{+31.23} & 59.23 & 70.75 & \textcolor{orange}{+11.52} \\
\midrule
\multicolumn{7}{l}{\textit{Math \& Reasoning Benchmarks}} \\
GSM8K~\citep{cobbe2021trainingverifierssolvemath}      & 80.82 & 86.96 & \textcolor{orange}{+6.14}  & 83.17 & 89.16 & \textcolor{orange}{+5.99} \\
MATH-500~\citep{hendrycksmath2021}       & 58.00 & 66.20 & \textcolor{orange}{+8.20}  & 60.40 & 72.60 & \textcolor{orange}{+12.20} \\
\midrule
\multicolumn{7}{l}{\textit{Science Benchmarks}} \\
SuperGPQA~\citep{pteam2025supergpqascalingllmevaluation}      & 33.85 & 39.62 & \textcolor{orange}{+5.77}  & 37.76 & 45.86 & \textcolor{orange}{+8.10} \\
SciBench~\citep{wang2024scibench}       & 18.50 & 23.30 & \textcolor{orange}{+4.80}  & 20.30 & 25.00 & \textcolor{orange}{+4.70} \\
\midrule
\multicolumn{7}{l}{\textit{General Capability Benchmarks}} \\
MMLU~\citep{hendrycks2021measuringmassivemultitasklanguage}           & 83.57 & 83.57 & {+0.00}                    & 86.37 & 87.37 & \textcolor{orange}{+1.00} \\
MMLU-Pro~\citep{wang2024mmluprorobustchallengingmultitask}       & 61.60 & 67.40 & \textcolor{orange}{+5.80}  & 63.80 & 72.70 & \textcolor{orange}{+8.90} \\
TriviaQA~\citep{joshi2017triviaqalargescaledistantly}       & 59.06 & 60.47 & \textcolor{orange}{+1.41}  & 74.29 & 77.14 & \textcolor{orange}{+2.85} \\
\bottomrule

\end{tabular}

\caption{General capability benchmarks comparing base and OpenSWE. $\Delta$ denotes absolute improvement.}
\label{tab:general_capability}
\end{table}

To assess whether SWE-focused training affects broader model capabilities, we evaluate OpenSWE models against their base counterparts on a suite of general benchmarks spanning code generation, mathematical reasoning, scientific knowledge, and general language understanding. Results are reported in Table~\ref{tab:general_capability}.

The largest gains appear on code benchmarks, where the 32B model improves by over 29 points on HumanEval and HumanEval+; because SWE tasks inherently require reading, editing, and generating code, this direct skill overlap yields the strongest transfer. Consistent improvements across all three math benchmarks suggest that the multi-step planning and logical decomposition cultivated by SWE debugging generalize to mathematical reasoning, even without explicit math training data. SuperGPQA and SciBench show moderate gains, likely because scientific questions demand structured inference chains similar to those practiced during patch generation, though the domain gap limits the magnitude. In contrast, MMLU remains nearly flat and TriviaQA improves only marginally, confirming that SWE training enhances procedural problem-solving capacity without affecting factual recall, which depends on pre-training coverage rather than reasoning ability.

\section{Conclusion}

We presented OpenSWE, the largest fully transparent framework for SWE agent training, comprising 45,320 executable Docker environments across 12.8k repositories with all infrastructure open-sourced. Through a multi-agent synthesis pipeline deployed on a 64-node cluster and a quality-centric filtering process addressing PR-Issue misalignment and triviality, we curated approximately 10,000 high-quality environments that provide a stronger training signal than existing alternatives.

Extensive experiments validate the effectiveness of OpenSWE: OpenSWE-32B and OpenSWE-72B achieve 62.4\% and 66.0\% on SWE-bench Verified, establishing state-of-the-art among SFT-based methods. Models trained on OpenSWE consistently outperform those trained on SWE-rebench across all model sizes and scaffolds, exhibit a log-linear data scaling trend with no observed saturation, and show improved general capabilities in code generation, mathematical reasoning, and scientific knowledge without degrading factual recall.

\bibliographystyle{acl_natbib}
\bibliography{bib}
\appendix

\section*{Appendix}
\section{SWE Environment Builder: Architecture and Prompt Excerpts}
\label{appendix:prompts}

This appendix documents the design of the \emph{builder} 
subsystem responsible for synthesizing reproducible Docker-based evaluation 
environments.

\textbf{Goal.} Given a task instance, which consists of a repository snapshot
at a fixed base-commit together with the patch information used for evaluation, 
our builder produces a Dockerfile that builds an isolated runtime 
environment and a bash evaluation script that runs the relevant tests while 
emitting machine-readable signals.

\textbf{Iterative loop.} The builder follows an iterative procedure. It first 
performs context retrieval by inspecting the repository to infer dependencies, 
Python constraints, and test entry points. It then synthesizes or retrieves, 
when available, a Dockerfile and an evaluation script. Then it executes and 
validates the resulting environment by building the image, running the 
evaluation script, and extracting structured markers from the logs. Finally, 
it refines the artifacts by providing concise failure diagnoses and repeating 
the loop.

\subsection{Prompt Design}
\label{appendix:prompt_design}

Below we quote only the prompt fragments that most directly enforce the
engineering invariants required for stable, large-scale synthesis.

\paragraph{Repo Exploration Agent.}
The retrieval prompt enforces a goal-driven and non-exhaustive policy. 
It discourages broad repository crawling and instead requires a short, 
actionable report that records exact versions and concrete test commands.

\begin{promptbox}{Repo exploration system prompt excerpt (verbatim)}
\scriptsize
\begin{verbatim}
You are a context_retrieval_agent responsible for gathering **precise and 
necessary information** from the local repository to support environment 
setup and test execution. After gathering the information, you will 
**generate a concise report** summarizing the key findings related to 
the setup and test execution.

Sometimes, another agent (such as a test analysis agent) may explicitly 
request specific information to help fix issues like Dockerfile errors or 
evaluation script failures.

Your primary goal is to:

- **If a specific request is provided by a calling agent, focus your 
  retrieval narrowly on that request, extracting only the explicitly 
  required files or data.**
- **If no explicit request is given by another agent, or if the request 
  is incomplete or unclear, perform a basic and limited exploration of 
  the repository to collect general environment and test execution 
  information. Avoid exhaustive or in-depth searches.**
- **Pay special attention to the following information when collecting 
  and summarizing:**
  - **Exact versions** of dependencies, libraries, and programming 
    languages (e.g., `flask==2.0.3`, `python3.9`, `node 18`)
  - **Commands** for setting up the environment and executing tests 
    (e.g., `pip install -r requirements.txt`, `pytest tests/test_api.py`)
  - Any environment configuration details (e.g., `.env` files, specific 
    OS package dependencies, etc.)
  - Specific test commands for individual or specific test files, not 
    just generic test execution commands.

### Suggested Retrieval Areas

Only investigate the following areas **if explicitly requested** by 
the calling agent. Focus your retrieval on the minimal set of files 
or configurations needed to resolve the issue efficiently and accurately.

1. **Environment Setup Information**
   - **Exact dependencies and their versions**: This includes dependencies 
     listed in files like `requirements.txt`, `pyproject.toml`, etc. 
     Ensure that the exact version for each dependency is captured.
   - **Programming language versions**: Ensure to capture version 
     information like Python (e.g., `python3.9`), and others as specified 
     in relevant configuration files (`.python-version`, etc.)
   - **Environment configuration files**: Collect details from `.env`, 
     `.bashrc`, or `.zshrc` if applicable, focusing on version-dependent
     environment variables and paths.
   - **OS-specific requirements**: Note any OS-dependent configurations 
     (e.g., specific Linux package dependencies in `apt` or `yum`).

2. **Test Execution Information**
   - **Precise test commands**: Focus on specific commands or instructions 
     for running individual tests or specific test files, not just commands 
     for running all tests. Look for test commands in documentation like 
     `README.md`, `CONTRIBUTING.md`, `tests/README.md`, etc.
   - **CI/CD configurations**: Look into files like `.github/workflows/`, 
     `.ci.yml`, `travis.yml`, or other pipeline configuration files that 
     might include commands for running tests or specific test environments.
   - **Test execution in context**: Extract any specific instructions about 
      running tests, such as flags for specific test cases, test suites, 
      or environments. Also, pay attention to dependencies relevant to 
      testing like test frameworks (e.g., `pytest`, `JUnit`, `Mocha`), 
      env frameworks (e.g. poetry, uv), and their versions.

3. **Organize Results for other agents**
   - Present findings in a structured way so they can be used to generate 
     the Dockerfile and evaluation script accurately. The **final report** 
     should:
     - Highlight the **specific versions** of dependencies, libraries, and 
       testing tools.
     - Include **commands** for setup and testing (e.g., `pip install`, 
       `npm install`, `pytest`).
     - Note any environment variables or configuration details relevant 
       to the environment setup and test execution.
     - Provide clear, concise, and actionable information, making it easier 
       for other agents to proceed with resolving any setup or test execution
       issues.

### Important Notes:
- The repository has already been **cloned locally**; you are working 
  within the local repository directory.  
- You are **not expected to search broadly**; retrieve only the files 
  and information explicitly requested by the calling agent.  
- Avoid redundant or speculative searches—**be goal-driven and 
  cost-efficient**.  

\end{verbatim}
\end{promptbox}

\paragraph{Dockerfile Agent.}
The Dockerfile prompt encodes hard constraints that prevent common 
failure modes, such as selecting an incorrect base image, omitting 
conda activation, or accidentally running tests during image construction. 
These constraints complement the architectural choices described in 
Section~\ref{dockerfile-construction} by making them \emph{non-negotiable} 
during generation.

\begin{promptbox}{Dockerfile init prompt excerpt (verbatim)}
\scriptsize
\begin{verbatim}
Generate a **Dockerfile** based on the collected environment setup information.
The Dockerfile must ensure that the provided test files can be executed
correctly.

### **Requirements:**
1. **Copy the repository** inside the Docker container into `/testbed/` and set
  `WORKDIR` to `/testbed/`.
2. **Checkout a specific commit SHA**, which will be provided by the user.
3. **Set up the environment** based on the information from the context
  retrieval agent:
   - Install necessary system dependencies and programming language versions.
   - Set up a virtual environment (`testbed`) if required.
   - Install all necessary libraries and dependencies.
4. **Ensure test execution** by setting up all necessary configurations.

### Important Notes:
1. You are FORBIDDEN to run tests in the dockerfile, tests will be run using
  eval script.
2. When building the Dockerfile, you MUST prioritize using package managers such
  as APT, Maven, or NPM etc to set up the environment efficiently.
3. Ensure shell compatibility by using `/bin/bash` as the default shell
  environment to avoid runtime issues.
4. Instead of using Ubuntu/Debian Docker image, You **MUST** directly use our
  provided `openswe-python-version` to setup python environment. It is built
  from

<dockerfile>
FROM continuumio/miniconda3:25.3.1-1
RUN sed -i 's|deb.debian.org|mirrors.cloud.aliyuncs.com|g'
/etc/apt/sources.list.d/debian.sources && \\
 apt update && \\
 rm -rf /var/lib/apt/lists/*
RUN conda create -n testbed python={python_version} -y; \\
 echo "conda activate testbed" >> ~/.bashrc; \\
 conda activate testbed; \\
pip config set global.index-url http://mirrors.cloud.aliyuncs.com/pypi/simple/; \\
 pip config set global.trusted-host mirrors.cloud.aliyuncs.com; \\
</dockerfile>
- It provides conda on debian 12, a python env named `testbed` with given
  version, and change mirror source.
- Available python versions include 2.7 and 3.5 to 3.14. Conda does not provide
  other versions. Chose **best fit version** rather than minimal.
- If a different base image is really necessary, please also change mirror to
  aliyun.
- It use a conda environment, so all python/pip related run must run with `bash
  -lc` or `. /opt/conda/etc/profile.d/conda.sh && conda activate testbed`
- If you are rewriting because of python version issue, you MUST NOT create new
  conda env; instead change base image version.
- Simply ignore conda update / pip update warning, unless it is root cause of
  error
5. It is recommended to use `COPY` to copy local files into the Docker
  container, and use of well-known basic image (python, miniforge), to avoid
  network stuff.
6. DO NOT run tests in the Dockerfile**.  
  - Do not include commands like `npm test`, `pytest`, or `mvn test`, or `python
    -m import xxx` in the Dockerfile.
  - Tests will be executed separately, and running them during the Docker build
    stage is an unnecessary overhead.
  - You can skip tests during environment setup because this is not your job.
7. If there is a reference Dockerfile, use it as a guideline.
8. Do not use ENTRYPOINT.
9. When setting up dependencies for the target repository (e.g., `torch 3.33`),
  **DO NOT** install the package directly from external registries (e.g., PyPI,
  NPM, Maven Central) using commands like `pip install <package>` (e.g., `pip
  install torch`).
  Instead, **you can install the repository itself in development mode** (`pip
  install -e .` for Python, `npm link` for Node.js, or `mvn install` for Java) to
  ensure that the local repository’s code is correctly referenced during
  execution.
   **Why is this important?**  
- If you modify the repository’s source code but have already installed a
  pre-built package from the registry, your system may load the installed package
  instead of your local code, **leading to incorrect test results and making
  debugging difficult**.
- Using development mode installation (`pip install -e .`, `npm link`, `mvn
  install`) ensures that the system always references the latest local repository
  code, preventing version mismatches and ensuring that modifications are properly
  reflected in subsequent tests.

### **Example Format:**
The Dockerfile must be wrapped in `<dockerfile>` tags. Example:

<dockerfile>
# Base image specification. Defines the foundation OS and python version for the
container (Required)
FROM openswe-python-3.12
# Fetch source code. same as git clone {{task.repo_name}} && git reset --hard
{{task.commit}} but avoid network stuff; guarantee to ready
COPY repo /testbed
# set default workdir to testbed. (Required)
WORKDIR /testbed/
# The lines above should NEVER change (except python version), so as to reuse
layers.

# Install package and environment manager required by the repo. (Example)
ENV DEBIAN_FRONTEND=noninteractive
RUN apt install -qq -y g++
# Target Project setup. Configures it, and installs project-specific
dependencies (Example)
# Note for conda, `-lc` is required for env activate; multicommand can split by
`;`
RUN bash -lc 'pip install -r requirements.txt' # install requirements from
context
RUN bash -lc 'pip install -e' # install self; important for running test
RUN bash -lc 'pip install pytest "poetry>=1,<2"' # special char need quote
</dockerfile>
\end{verbatim}
\end{promptbox}

\paragraph{Write Evaluation Script Agent.}
The evaluation-script prompt enforces a deterministic and judge-friendly 
interface. It requires non-interactive patch application via heredoc 
placeholders and mandates that test execution emit machine-readable 
markers to support rule-based extraction.

\begin{promptbox}{eval-script init prompt excerpt (verbatim)}
\scriptsize
\begin{verbatim}
Generate an **evaluation script** based on the collected environment setup and
test execution information.
The script must execute the provided test files inside the specified Docker
environment.

### **Requirements:**
1. **Activate the environment**: Ensure the correct environment (e.g., Conda,
  venv) is activated before running the tests.
2. **Apply the patch**: The patch may need to be applied before running the
  tests.
3. **Execute the given test files and unittests** using the correct command
  found by the context retrieval agent.

### Important Notes:
1. You must **execute only the specified target test files and unittests**,
rather than running all tests in the repository.
  - Running all tests can be highly time-consuming and unnecessary.  
  - Ensure that only the **required test files** are executed. You may refer to
    golden patch, but please remain some already passed tests other than fixed in
    golden patch.

2. **Optimize execution efficiency by combining multiple test commands into a
  single command** whenever possible.
  - Avoid running multiple separate test commands if they can be executed in one
    batch.
  - This reduces redundant initialization overhead and speeds up execution.  

3. **Ensure that the output of the evaluation script is concise and
  structured**, making it easier for the **test log analysis agent** to process.
  - The test command **must output the names and pass/fail/skip status of each
    target executed test file**.
  - Avoid excessive debug information or unrelated output in eval script, but **do
    not suppress key test execution details**.
  - Avoid running all tests! **Just run the target unittests fixed by gold
    patch**.

4. **Follow the structure of the reference evaluation script or eval script
  skeleton whenever available.
  - Use **a simple, minimalistic structure** similar to the reference eval script
    to ensure clarity and maintainability.
  - The script should be easy to modify and extend without unnecessary complexity.

5. **The actual test patch content is omitted here for brevity (marked with
  [CONTENT OF TEST PATCH] placeholder).
  - You must generate the complete git apply command structure, including the
    heredoc syntax with delimiter (EOF_114329324912).
  - The placeholder will be programmatically replaced with the actual patch content
    during script execution.
    - Example structure:
      git apply -v - <<'EOF_114329324912'\n
      [CONTENT OF TEST PATCH]\nEOF_114329324912

6. You MUST capture the exit code immediately after running the tests using
  `rc=$?`, and then echo: `OPENSWE_EXIT_CODE=$rc`. This ensures the judge can
  determine whether the tests passed successfully. Also, you MUST NOT include `set
  -e`, which will truncate out error code.

7. You MUST print ">>>>> Start Test Output" exactly before test (pytest for
  example), and ">>>>> End Test Output" after it, we will extract output with it
  after run.

### **Example Format:**
The script must be wrapped in `<script>` tags. Example:

<script>
#!/bin/bash
# activate environment
. /opt/conda/etc/profile.d/conda.sh
conda activate testbed # already created by base image
cd /testbed

# Required: apply test patch to update target tests
git apply -v --allow-empty - <<'EOF_114329324912'
[CONTENT OF TEST PATCH]
EOF_114329324912

# Required: run target tests files instead of all tests!
echo ">>>>> Start Test Output"
pytest --no-header -rA --tb=no -p no:cacheprovider -n4
mypy/test/testcheck.py::TypeCheckSuite::check-functions.test
mypy/test/testcheck.py::TypeCheckSuite::check-redefine.test
rc=$? # Required, save exit code
echo ">>>>> End Test Output"
echo "OPENSWE_EXIT_CODE=$rc" #Required, echo test status
</script>
\end{verbatim}
\end{promptbox}

\paragraph{Test Analysis Agent.}
The analysis prompt turns verbose logs into actionable iteration signals by enforcing a rule-based validity criterion, under which the test-only run must fail while the run with the fix must pass. It also specifies explicit routing: when a failure is attributed to the Dockerfile rather than the evaluation script, the feedback is directed to the corresponding writer agent.

\begin{promptbox}{Test analysis prompt excerpt (verbatim)}
\scriptsize
\begin{verbatim}
Given the test log and the target tests, analyze the results and determine the
next steps. But if the dockerfile is not built successfully, you should analyze
what issues happen.

### **Step 1: Verify Test Execution**
- Identify which test files were added or modified by the eval script.
- Confirm that those tests were actually executed (they appear in the test log).
- Check their return code:
   - Return code for testOnly MUST BE non-0 and for testWithFix MUST BE 0.
   - Check their pass/fail status:
   - If all tests switch from fail to pass, report success.
- If there exists fail to fail or pass to fail, report fail. (MUST fix by write
  eval agent)
- If there exists pass to pass, and every other thing is correct, you may report
  success.
- Ensure there is at least some test output in the log:
- If no test output is found, set `is_finish = false` and include an instruction
  for write_eval_script_agent to revise the eval script so that tests actually
  run.

### **Step 2: Identify Problems**
- If the tests failed due to **environment setup issues**, analyze whether the
  problem comes from:
- The **Dockerfile** (e.g., incorrect dependencies, wrong OS, missing
  configurations).
- The **evaluation script** (e.g., incorrect test commands, wrong paths, missing
  environment activation, mismatch with unit tests solved by the gold patch).
- Simply ignore conda update / pip update warning, unless it is root cause of
  error
- Sometimes, tests may fail due to incorrect versions of specific dependencies.
  Be sure to check the versions of critical dependencies to ensure compatibility.
- If there are missing dependencies or unknown errors, consider whether
  additional context retrieval is required.
- Tests should not be run in the Dockerfile**; skip tests during environment
  setup and run them in the evaluation script.
- Note that the eval script MUST catch exit code after running tests, and echo
  "OPENSWE_EXIT_CODE=$rc". This is important for judge whether tests are run
  successfully.

### **Step 3: Plan Corrective Actions**
- If a fix is needed in the **Dockerfile**, provide guidance to
  `write_dockerfile_agent` on how to fix it, always include the original error
  message and a brief description of what is missing or suspected to be the cause.
- If a fix is needed in the **evaluation script**, provide guidance to
  `write_eval_script_agent` on how to fix it, always include the original error
  message and a brief description of what is missing or suspected to be the cause.
- If more information from the target repository is needed, provide guidance to
  `context_retrieval_agent` on what to collect. Here are some instructions:
  1. Always include the original error message and a brief description of what is
    missing or suspected to be the cause.
  2. Clearly specify what information or files should be searched for. For
    environment or dependency issues, recommend files such as requirements.txt,
    environment.yml, Dockerfile, setup.py, pyproject.toml, etc. For test or
    evaluation issues, suggest looking for files such as  eval*.sh, pytest.ini,
    .github/workflows/*, etc.
  3. Additionally, encourage reviewing documentation files like README.md,
    CONTRIBUTING.md, or any docs in the root or docs/ directory for relevant setup
    or testing instructions (Contributing file often contains some testing
    instruction).
  4. Always add guidance to at least one of dockerfile agent or eval script agent
    if you guide to context retrival agent, otherwise nothing is rewritten and 
    error will replay.
    - If you encounter network issue, simply put all guidance empty and set
      is_finish to false; we will rerun it.
    - If you think the issue is unsolvable, you may simply set is_finish to true,
      sparing effort; for example:
      1. Golden patch does not solve any unittest.
      2. Dependency of project has unsolvable conflicts
3. Some dependency have become missing, like 404 file, super old versions (numpy
  <= 1.8)...

### **Output Example**
Provide your answer in JSON format:
```json
{
"is_finish": true/false,  # If tests passed and everything is correct or the
issue is considered unsolvable, set this to true.
"guidance_for_write_dockerfile_agent": "<Provide detailed guidance if
modifications are needed>",
"guidance_for_write_eval_script_agent": "<Provide detailed guidance if
modifications are needed>",
"guidance_for_context_retrieval_agent": "<Specify what additional information
from the target repository is needed, if any>",
}
```

**Important Notes:**
- If `is_finish` is `true`, all guidance fields can be empty.
- Be specific in your guidance, providing detailed steps for the necessary
  fixes. Only provide guidance to the relevant agent based on the actual issue.
  For any agent not called, its guidance field must be empty.
- Calling context_retrieval_agent is expensive. Only suggest using it when there
  is clearly missing information that is necessary to fix the Dockerfile or
  evaluation script. Be precise and specific in what to retrieve (e.g., particular
  files or configuration scripts) to avoid repeated or vague searches.
- Provide detailed error information to tell agent what errors happen.

\end{verbatim}
\end{promptbox}

\section{Construction Cost Estimate}
\label{appendix:cost_estimate}

Based on the 64-node configuration in Table~\ref{tab:hardware_config}, we provide
an approximate 10-day construction cost estimate in Table~\ref{tab:cost_estimate}.
The total construction budget is primarily sensitive to effective
GPU-hour price and cluster utilization efficiency. In practice, preemptible
pricing, committed-use discounts, and scheduling efficiency can substantially
change the final amount.

\paragraph{Curation Cost.} Beyond environment construction, the trajectory sampling and difficulty-aware curation process requires an additional computational investment of approximately \$576,000. This cost primarily comprises LLM API expenses for generating resolution trajectories using the GLM-4.7 model across the full OpenSWE and SWE-rebench datasets (four attempts per instance under the OpenHands and SWE-Agent scaffolds), as well as the associated Docker compute costs for executing each trajectory within its corresponding environment. Combined with the environment construction budget, the total cost of the OpenSWE project exceeds \$1.47 million.

\begin{table}[h]
\centering
\begin{tabular}{r c c}
\hline
\textbf{Cost Item} & \textbf{Estimated Cost (USD)} & \textbf{Cost per Instance (USD)} \\ \hline
Storage & \$13,000 & \$0.29 \\
CPU & \$7,000 & \$0.15 \\
Network & \$3,000 & \$0.07 \\
Container Registry Service & \$3,000 & \$0.07 \\
GPU  & \$865,000 & \$19.08 \\ \hline
\textbf{Total} & \textbf{\$891,000} & \textbf{\$19.66} \\ \hline
\end{tabular}
\caption{Approximate construction cost for a 64-node, 10-day run.}
\label{tab:cost_estimate}
\end{table}

\end{document}